\documentclass[fleqn,useAMS,usenatbib]{mnras}
\usepackage{epsfig,graphicx}
\usepackage{graphics}
\usepackage{amsmath,amssymb,txfonts}
\usepackage{xspace,ulem}

\newcommand{\xte}{{\textit{RXTE}}\xspace}
\newcommand{\chandra}{{\textit{Chandra}}\xspace}

\newcommand{\swift}{{\textit{Swift}}\xspace}
\newcommand{\msun}{{\rm M}_{\sun}}

\begin{document}
\title[Correlations between radio and bolometric fluxes]{Correlations between radio and bolometric fluxes in GX~339--4 and H1743--322}
\date{}
\author[N. Islam and A. A. Zdziarski]{Nazma Islam\thanks{E-mail: nislam@camk.edu.pl (NI), aaz@camk.edu.pl (AAZ)} and Andrzej A. Zdziarski\footnotemark[1]\\
Nicolaus Copernicus Astronomical Center, Polish Academy of Sciences, Bartycka 18, 00-716 Warsaw, Poland}

\date{Accepted 2018 September 14, Received 2018 September 14; in original form 2018 July 18}

\pagerange{\pageref{firstpage}--\pageref{lastpage}}
\pubyear{2018}
\maketitle
\label{firstpage}

\begin{abstract}
Compact radio jets are ubiquitous in stellar-mass black-hole binaries in their hard spectral state. Empirical relations between the radio and narrow-band X-ray fluxes have been used to understand the connection between their accretion discs and jets. However, a narrow-band (e.g., 1--10 or 3--9 keV) X-ray flux can be a poor proxy for either the bolometric luminosity or the mass accretion rate. Here, we study correlations between the radio and unabsorbed broad-band X-ray fluxes, the latter providing good estimates of the bolometric flux. We consider GX~339--4, the benchmark object for the main branch of the correlation, and H1743--322, the first source found to be an outlier of the correlation. The obtained power-law dependencies of the radio flux on the bolometric flux have significantly different indices from those found for the narrow X-ray bands. Also, the radio/bolometric flux correlations for the rise of the outbursts are found to be significantly different from those for the outburst decline. This points to a possible existence of a jet hysteresis in the radio/X-ray source evolution, in addition to that seen in the hardness/flux diagram of low-mass X-ray binaries. The correlation during the rise of the outbursts is similar for both GX~339--4 and H1743--322. The correlation for the decline of the outbursts for H1743--322 lies below that of GX~339--4 at intermediate X-ray fluxes, whereas it approaches the standard correlation at lower X-ray luminosities. We also compare these correlations to those for the high-mass X-ray binaries Cyg~X-1 and Cyg~X-3. 
\end{abstract}

\section{Introduction}

Stellar mass black-hole (BH) X-ray binaries (BHXRBs) display a wide range of emission properties, pertaining to its X-ray spectral state \citep{remillard2006}. A majority of the BHXRBs are transient sources, spending most of the time in a quiescent state. Occasionally they go into outbursts, during which the X-ray fluxes increase by several orders of magnitude. The two main spectral states seen in these outbursts are a high soft state, dominated by thermal emission from an accretion disc, and a low hard state, which spectra are dominated by Comptonization with a high-energy cutoff above $\sim$100 keV, and a weak disc component. 

A salient feature of the BHXRBs in the low hard state is the presence of compact jets \citep*{fender2004,fender2009}. Partially synchrotron self-absorbed emission from the jets extends from radio to at least infrared (e.g. \citealt{blandford1979, hjellming1995,russell2006,russell2013, corbel2013a, corbel2013b}). On the other hand, the X-ray emission is most likely dominated by the accretion flows. Relationships between the radio and X-ray emission have been used as tools to study the nature of the accretion disc-jet coupling in these systems.

Simultaneous multi-wavelength observations of various BH binaries in their outbursts have provided evidence of a strong non-linear correlation between the radio flux, $F_{\rm R}$, and the X-ray flux, $F_{\rm X}$ (\citealt{hannikainen1998,corbel2003,corbel2013a}; \citealt*{gallo2003}; \citealt{gallo2006}; \citealt*{gallo2012,gallo2018}; \citealt{rushton2016}). The correlation has an approximate power-law form, $F_{\rm R} \propto F_{\rm X}^{b}$. Most of the BH sources lie on the so-called standard track, with the power-law index of $b \sim 0.5$--0.7, which has been interpreted as due to the X-rays being emitted by radiatively inefficient accretion flows \citep{corbel2003,gallo2003,gallo2012,gallo2014}. However, there is a number of outlier sources, which are found to lie well outside the standard correlation, and show a steeper power-law index of $b \sim 1.4$, which can be interpreted as due to the presence of radiatively efficient accretion flows \citep{bel2007,soleri2010,soleri2011,coriat2011,jonker2012,ratti2012,russell2015,plotkin2017}. The correlations demonstrate the existence of strong and relatively stable couplings between the radio emitting jets and the X-ray emitting accretion flows, over a wide range of luminosities and for different outbursts. These radio/X-ray correlations are crucial for defining the so-called fundamental plane of BH activity \citep*{merloni2003,falcke2004}. Fig.\ 9 in \cite{corbel2013a} shows the radio versus X-ray luminosities of several BHXRBs in their hard and quiescent states, exhibiting the standard and outlier tracks. These radio/X-ray correlations are dominated by two sources, GX 339--4 for the standard track, and H1743--322 for the outlier track. 

However, those studies have utilised the X-ray flux in a narrow energy band of either 1--10 keV or 3--9 keV. This narrow-band flux has been used as a proxy for the accretion rate, which can bear a substantial error given most of the luminosity in the hard state is emitted around $\sim$100 keV. This problem was pointed out by \citet{zdziarski2004}, who calculated a radio vs.\ bolometric flux correlation for a sample of observations of GX 339--4. Then, \citet{zdziarski2011} and \citet*{zdziarski2016} performed analogous calculations for Cyg X-1 and Cyg X-3, respectively. In both of the sources, the power-law indices of narrow X-ray bands were found to strongly depend on the choice of the energy range. In fact, the correlation for Cyg X-3 changes sign from positive at soft X-rays to negative at hard X-rays. These results strongly argue for the use of the bolometric flux in studies of the radio/X-ray correlation. In the present paper, we perform such a study for the two archetypal sources, GX 339--4 and H1743--322. Both of them underwent multiple outbursts, and have simultaneous coverage in radio and X-rays.

The low-mass BHXRB GX 339--4 was discovered in 1972 by the {\it OSO-7\/} satellite \citep{markert73}. It has since then been the most often outbursting transient BHXRB. Its distance appears to be around $D \sim 8$ kpc \citep{zdziarski2004}, though it is not well constrained. \citet{heida2017} found a preferred $D\sim 9$ kpc and a lower limit of $\sim$5 kpc.  Its mass function has been measured by \citet{heida2017}, implying, with some other constraints, $M_{\rm BH} < 9.5\msun$. When calculating the luminosity and Eddington ratio for the source, we assume $M_{\rm BH} = 6\msun$ (which is a preferred value of \citealt{heida2017}) and $D=8$ kpc.

The X-ray transient H1743--322 was discovered during an outburst in 1977 with the {\it Ariel V\/} and {\it HEAO-1\/} satellites \citep{kaluzienski1977}. Currently there are no dynamical estimates on its BH mass. Its spectral and timing features were found to be similar to those seen in the BHXRB XTE J1550--564 \citep{mcclintock2009}, suggesting a similar BH mass, which in XTE J1550--564 is within 8--$14\msun$ \citep{orosz11}. The distance to H1743--322 has been estimated as $8.5\pm 0.8$ kpc \citep*{steiner12}, compatible with its location near the Galactic Centre, from which its projected separation is only $\simeq\! 2\degr$. When calculating the luminosity and Eddington ratio for the source, we assume $M_{\rm BH} = 10\msun$ and $D=8.5$ kpc.

In Section \ref{data}, we describe the observations and our modelling of the X-ray spectral data. We present our results in Section \ref{results} and conclusions in Section \ref{conclusions}.

\begin{table}
\caption{The list of the GX 339--4 fluxes used in this work. The radio fluxes, $F_{\rm R}$, at either 8.6 or 9.0 GHz are from \citet{corbel2013a} and the bolometric fluxes, $F_{\rm bol}$, have been calculated using the \xte observations within a day from a radio pointing, see Section \ref{obs_gx339}. \label{GX}}
\centering
\scriptsize
\begin{tabular}{c c c c}
\hline

Date      &     MJD	   &    $F_{\rm R}(8.6$ or $9.0$\,GHz)   & $F_{\rm bol}$    \\
 &   &     [mJy]&  [10$^{-9}$ erg cm$^{-2}$ s$^{-1}$]  \\
\hline
1997 February 04 &   50483      &    9.1 $\pm$  0.1  &  7.9  $\pm$  0.5    \\
1997 February 10 &   50489      &    8.2 $\pm$  0.2  &  7.3  $\pm$  0.5    \\
1997 February 17 &   50496      &    8.7 $\pm$  0.2  &  6.9  $\pm$  0.5     \\
1999 February 12 &   51221      &    4.6 $\pm$  0.08 &  3.0  $\pm$  0.1  \\
1999 March 3 &  51240      &    5.74 $\pm$ 0.06 &  3.7 $\pm$ 0.3  \\
1999 April 2 &  51270      &    5.1 $\pm$  0.06 &  4.2 $\pm$ 0.3   \\
1999 April 22 & 51290      &    3.2 $\pm$  0.06 &  2.0 $\pm$ 0.2  \\
1999 May 14  &  51312      &    1.44 $\pm$ 0.06 &  0.62 $\pm$ 0.04  \\
1999 June 25 &  51354      &    0.24 $\pm$ 0.05 &  0.04 $\pm$ 0.01 \\
1999 July 07 &  51366      &    0.12 $\pm$ 0.04 &  0.03 $\pm$ 0.01 \\
1999 August  17 &  51407      &    0.27 $\pm$ 0.07 &  0.014 $\pm$ 0.008   \\
2002 April 4 &  52368      &    5.95 $\pm$ 0.15 &  13.3 $\pm$ 0.4        \\	  
2002 April 7 &  52371      &    8.27 $\pm$ 0.07 &  21.7 $\pm$ 0.7 \\
2002 April 18 & 52382      &    14.27 $\pm$ 0.05 &  34.0 $\pm$ 1.0         \\
2003 May 25  &  52784      &    0.77 $\pm$ 0.06 &  0.13 $\pm$ 0.02   \\
2004 February 13  &  53048      &    1.13 $\pm$ 0.08 &  0.9 $\pm$ 0.2    \\
2004 February 24  &  53059      &    1.84 $\pm$ 0.2  &  2.9 $\pm$ 0.3    \\
2004 March 16 & 53080      &    4.88 $\pm$ 0.06 &  9.3 $\pm$ 0.6     \\
2004 March 17 & 53081      &    4.84 $\pm$ 0.11 &  9.8 $\pm$ 0.8     \\
2004 March 18 & 53082      &    4.98 $\pm$ 0.11 &  10.4 $\pm$ 0.8     \\
2004 March 19 & 53083      &    5.2 $\pm$  0.1  &  10.5 $\pm$ 0.8     \\
2005 April 21 & 53481      &    4.73 $\pm$ 0.05 &  3.4 $\pm$ 0.2   \\
2005 April 24 & 53484      &    4.23 $\pm$ 0.08 &  2.6  $\pm$ 0.2   \\
2005 April 28 & 53488      &    3.46 $\pm$ 0.13 &  1.9 $\pm$ 0.1   \\
2005 April 29 & 53489      &    3.32 $\pm$ 0.1  &  1.7 $\pm$ 0.1   \\
2005 April 30 & 53490      &    2.94 $\pm$ 0.07 &  1.6 $\pm$ 0.2   \\
2005 May 3   &  53493      &    1.92 $\pm$ 0.14 &  1.3 $\pm$ 0.1  \\
2005 May 4   &  53494      &    1.99 $\pm$ 0.1  &  1.2 $\pm$ 0.1   \\
2005 May 6   &  53496      &    1.69 $\pm$ 0.12 &  1.0 $\pm$ 0.1  \\
2005 May 12  &  53502      &    1.00 $\pm$ 0.18 &  0.7 $\pm$ 0.1   \\
2007 February 4   &  54135      &    22.5 $\pm$ 0.3  &  34.0  $\pm$ 1.0 \\
2007 May 31  &  54251      &    4.37 $\pm$ 0.09 &  1.9  $\pm$ 0.2         \\
2007 June 6  &  54257      &    2.63 $\pm$ 0.18 &  1.5  $\pm$ 0.1 \\
2007 June 11 &  54262      &    2.01 $\pm$ 0.15 &  1.2 $\pm$ 0.1 \\
2007 June 25 &  54276      &    1.69 $\pm$ 0.05 &  1.3 $\pm$ 0.2         \\
2007 June 29 &  54280      &    2.0  $\pm$ 0.2  &  1.5 $\pm$ 0.2 \\
2007 July 4  &  54285      &    2.1  $\pm$ 0.2  &  2.0 $\pm$ 0.2 \\
2007 July 13 &  54294      &    2.66 $\pm$ 0.05 & 2.9 $\pm$ 0.4 \\
2007 August  22 &  54334      &    2.95 $\pm$ 0.07 &  4.1 $\pm$ 0.5        \\
2007 November  3  &  54407      &    0.80 $\pm$ 0.07 &  0.2 $\pm$ 0.03    \\
2007 November  27 &  54431      &    0.48 $\pm$ 0.07 &  0.15 $\pm$ 0.03    \\
2008 June  26 &  54643      &    1.16 $\pm$ 0.10 &  0.92 $\pm$ 0.08       \\
2008 July 5  &  54652      &    1.24 $\pm$ 0.07 &  1.0 $\pm$ 0.1       \\
2008 July 16 &  54663      &    1.51 $\pm$ 0.06 &  1.5  $\pm$ 0.2       \\
2008 August  18 &  54696      &    1.18 $\pm$ 0.10 &  1.1 $\pm$ 0.1         \\
2008 October  10 &  54749      &    0.73 $\pm$ 0.10 &  0.19 $\pm$ 0.06     \\
2010 January  21 &  55217      &    5.05 $\pm$ 0.05 &  7.3 $\pm$ 0.8   \\
2010 February 13  &  55240      &    5.90 $\pm$ 0.10 &  9.1 $\pm$ 0.9        \\
2010 March 3 &  55258      &    7.30 $\pm$ 0.10 &  13.0 $\pm$ 1.0       \\
2010 March 6 &  55261      &    9.60 $\pm$ 0.05 &  15.0 $\pm$ 1.0        \\
2010 March 7 &  55262      &    8.05 $\pm$ 0.10 &  15.0 $\pm$ 1.0         \\
2010 March 14 & 55269      &    11.32 $\pm$ 0.10 &  17.0 $\pm$ 1.0         \\
2010 March 16 & 55271      &    12.04 $\pm$ 0.10 &  19.0 $\pm$ 1.0         \\
2010 March 24 & 55279      &    18.59 $\pm$ 0.05 &  25.0 $\pm$ 2.0       \\
2010 March 31 & 55286      &    25.94 $\pm$ 0.05 &  31.0 $\pm$ 2.0         \\
2010 April 2 &  55288      &    25.18 $\pm$ 0.10 &  30.0 $\pm$ 2.0        \\
2010 April 3 &  55289      &    21.11 $\pm$ 0.15 &  32.0 $\pm$ 2.0   \\
2010 April 4 &  55290      &    23.53 $\pm$ 0.05 &  33.0 $\pm$ 2.0         \\
2010 April 5 &  55291      &    24.69 $\pm$ 0.05 &  34.0 $\pm$ 2.0     \\
2010 April 6 &  55292      &    23.90 $\pm$ 0.06 &  35.0 $\pm$ 2.0     \\
2011 February   13 & 55605      &    4.17  $\pm$ 0.05 &  2.3 $\pm$ 0.1 \\
2011 February   15 & 55607      &    3.87  $\pm$ 0.05 &  2.2 $\pm$ 0.2    \\
2011 February   18 & 55610      &    3.98 $\pm$ 0.1  &  1.9 $\pm$ 0.2     \\
2011 February   20 & 55612      &    3.84 $\pm$ 0.05 &  1.3 $\pm$ 0.1    \\
2011 February   24 & 55616      &    2.95 $\pm$ 0.05 &  0.8 $\pm$ 0.1    \\
2011 February   27 & 55619      &    2.42 $\pm$ 0.08 &  0.6 $\pm$ 0.1    \\
2011 March 3  & 55623      &    1.64 $\pm$ 0.05 &  0.5 $\pm$ 0.1     \\
2011 March 7  & 55627      &    1.26 $\pm$ 0.1  &  0.5 $\pm$ 0.2   \\
2011 March 9  & 55629      &    1.38 $\pm$ 0.08 &  0.4 $\pm$ 0.1     \\
2011 March 20 & 55640      &    0.74 $\pm$ 0.04 &  0.17 $\pm$ 0.03   \\
\hline
\end{tabular}
\end{table}

\begin{table}
\caption{The list of the H1743--322 fluxes used in this work. The radio fluxes are from \citet{coriat2011} and the bolometric fluxes have been calculated using the \xte observations within a day from a radio pointing, see Section \ref{obs_gx339}. \label{H}}
\scriptsize
\centering
\begin{tabular}{c c c c c}
\hline
Date      &     MJD	   &    $F_{\rm R}(8.5$\,GHz)    & $F_{\rm bol}$    \\
 &   &     [mJy] &  [$10^{-9}$ erg cm$^{-2}$ s$^{-1}$ \\
\hline
2003 March 28  & 52728     &    4.57 $\pm$ 0.12      &  5.3 $\pm$ 0.1  \\
2003 April 1   & 52730     &    6.45 $\pm$ 0.12      &  25 $\pm$ 2   \\
2003 April 4   & 52733     &    21.81 $\pm$ 0.13     &  16 $\pm$ 1       \\
2003 April 6   & 52735     &    19.43 $\pm$ 0.16     &  21 $\pm$ 2 \\
2003 November 5     & 52948     &    0.22 $\pm$ 0.05      &  0.83 $\pm$ 0.05 \\
2004 November 1     & 53310     &    0.31 $\pm$ 0.06      &  0.13 $\pm$ 0.01  \\
2008 January 28    & 54493     &    0.44 $\pm$ 0.09      &  4.0 $\pm$ 0.1 \\
2008 February 3     & 54499     &    0.52 $\pm$ 0.07      &  2.7 $\pm$ 0.1  \\
2008 February 5     & 54501     &    0.48 $\pm$ 0.08      &  2.3 $\pm$ 0.1    \\
2008 February 6     & 54502     &    0.45 $\pm$ 0.09      &  2.0 $\pm$ 0.1     \\
2008 February 9     & 54505     &    0.56 $\pm$ 0.05      &  1.7 $\pm$ 0.1     \\
2008 February 19    & 54515     &    0.23 $\pm$ 0.12      &  0.22 $\pm$ 0.02   \\
2008 February 20    & 54516     &    0.21 $\pm$ 0.05      &  0.14 $\pm$ 0.03   \\
2008 February 23    & 54519     &    0.23 $\pm$ 0.06      &  0.13 $\pm$ 0.04   \\
2008 February 24    & 54520     &    0.31 $\pm$ 0.05      &  0.012 $\pm$ 0.004         \\
2008 October 5 & 54744     &    1.74 $\pm$ 0.07      &  8.0 $\pm$ 0.4   \\
2008 October 8 & 54747     &    2.54 $\pm$ 0.08      &  8.9 $\pm$ 0.3   \\
2008 October 9 & 54748     &    2.43 $\pm$ 0.09      &  8.1 $\pm$ 0.3    \\
2008 October 10 & 54749    &    2.38 $\pm$ 0.11      &  7.8 $\pm$ 0.4     \\
2008 November 4     & 54774     &    0.94 $\pm$ 0.12      &  4.2 $\pm$ 0.3        \\
2008 November 9     & 54779     &    0.94 $\pm$ 0.08      &  4.2 $\pm$ 0.3        \\
2009 May 30    & 54981     &    2.73 $\pm$ 0.1       &  8.8 $\pm$ 0.3         \\
2009 July 7    & 55019     &    0.592 $\pm$ 0.055    &  2.8 $\pm$ 0.1        \\
2009 July 9    & 55021     &    0.41 $\pm$ 0.074     &  2.5 $\pm$ 0.1 \\
2009 July 12   & 55024     &    0.335 $\pm$ 0.063    &  2.2 $\pm$ 0.1       \\
2009 July 13   & 55025     &    0.587 $\pm$ 0.066    &  1.8 $\pm$ 0.1        \\
2009 July 19   & 55031     &    0.631 $\pm$ 0.052    &  1.2 $\pm$ 0.1 \\
2010 February 13    & 55240     &    0.28 $\pm$ 0.05      &  1.0 $\pm$ 0.3 \\
\hline
\end{tabular}
\end{table}

\section{Observations and Data Analysis}
\label{data}

\begin{figure*}  
\includegraphics[height=\textwidth,angle=-90]{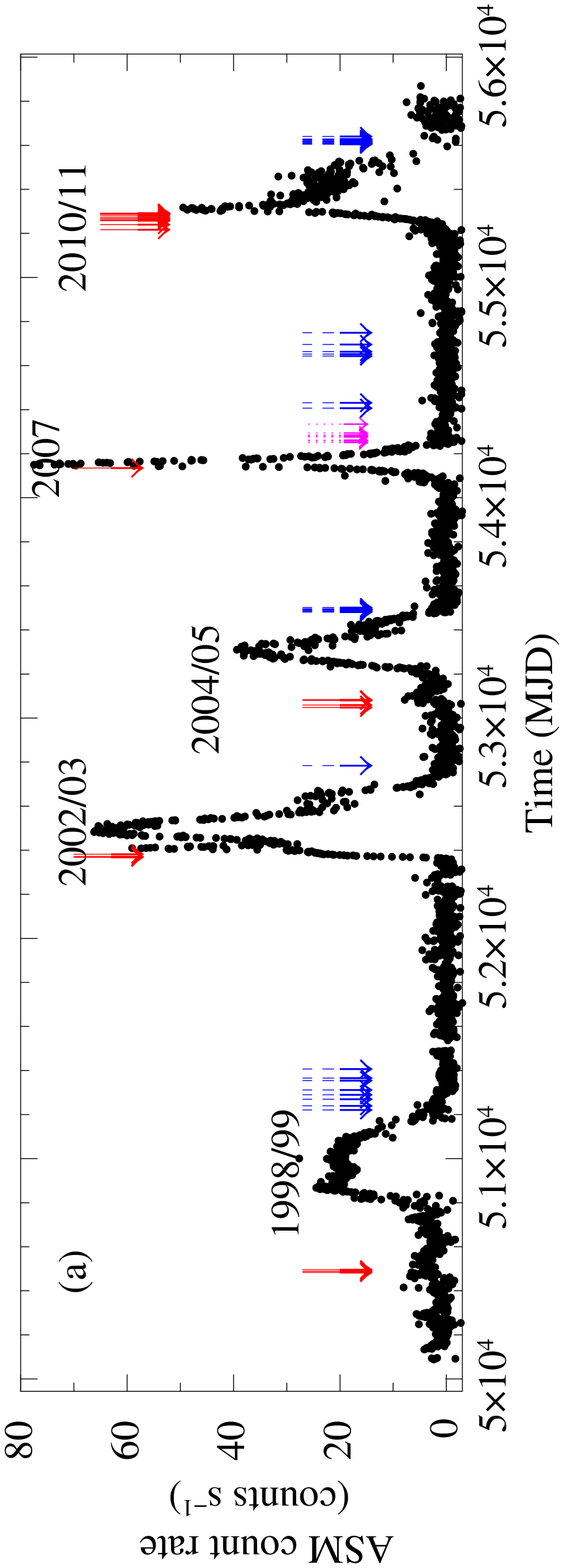}      
\includegraphics[height=\textwidth,angle=-90]{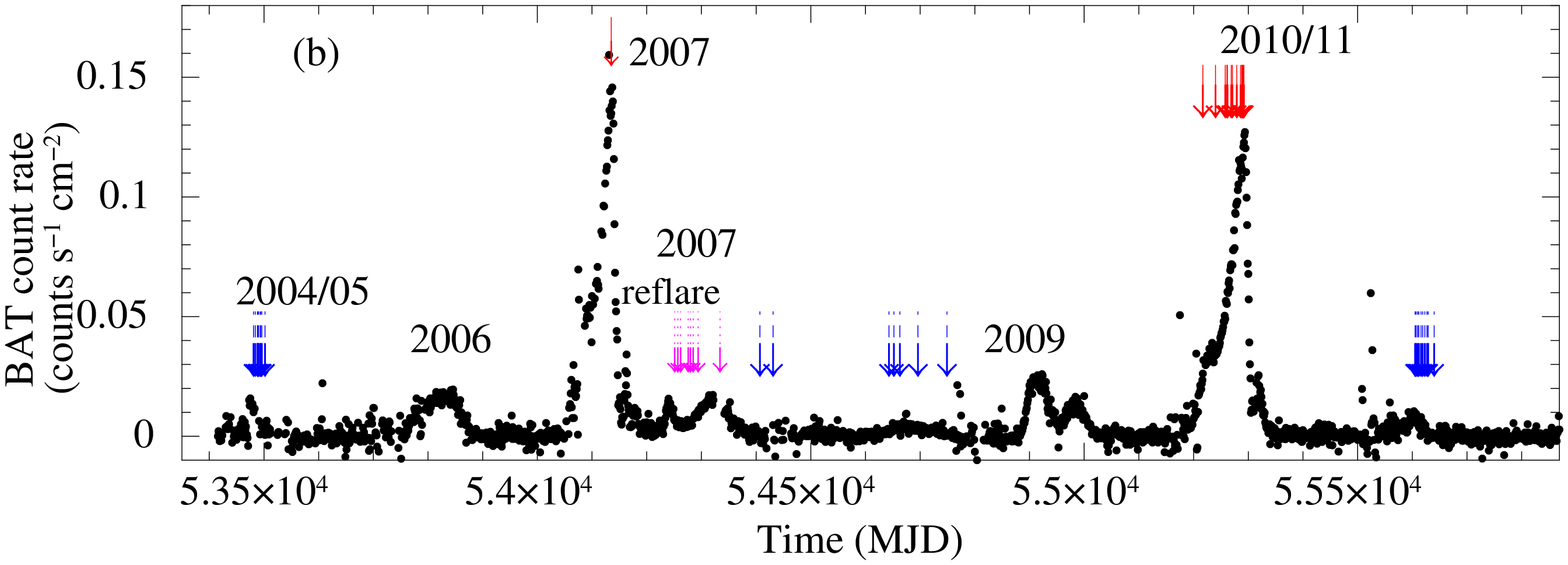}
\caption{The (a) ASM (1.5--12 keV) and (b) BAT (15--50 keV) daily light curves for GX 339--4 during the \xte lifetime. As seen in (a), the source underwent five major outbursts during that time, in 1998/99, 2002/03, 2004/05, 2007 and 2010/11. The BAT light curve shows in addition hard outbursts in 2006 and 2009, as well as that in 2007, which may be a reflare of the 2007 outburst. The red solid and blue dashed arrows mark the pointed \xte observations carried out in the hard state during its rise and decline, respectively. The magenta dotted arrows mark the 2007 reflare, which assignment to either rise or decline appears uncertain. The error bars are not plotted for clarity, here and in Fig.\ \ref{h1743_lc}.}
\label{lc}
\end{figure*}

\begin{figure*}     \includegraphics[height=\textwidth,angle=-90]{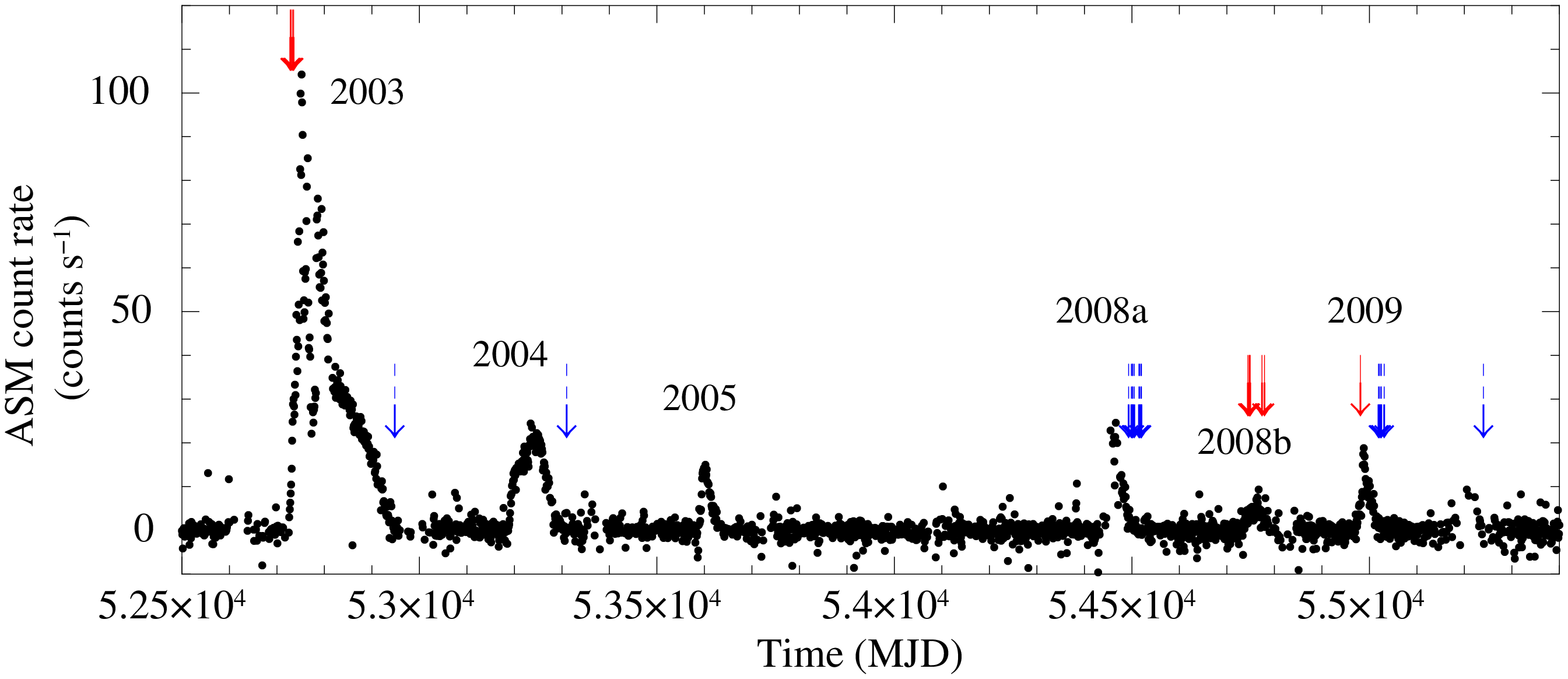}   
\caption{The 2003--2010 ASM (1.5--12 keV) light curve from for H1743--322. The source underwent a major outburst in 2003, followed by five minor outbursts, marked as 2004, 2005, 2008a, 2008b and 2009. The red solid and blue dashed arrows mark the pointed \xte observations carried out in the hard state during its rise and decline, respectively.}
\label{h1743_lc}
\end{figure*}

\subsection{X-ray light curves of GX 339--4 and H1743--322}
\label{light_curves}

{\it Rossi X-ray Timing Explorer}\/ (\xte) was an X-ray observatory operational from 1996 to 2012. The Proportional Counter Array (PCA; \citealt{jahoda2006}) and the High Energy X-ray Timing Experiment (HEXTE; \citealt{rothschild1998}) were the pointed detectors on board \xte. The PCA and HEXTE operated in the nominal energy ranges of 2--60 and 15--250 keV, respectively. \xte conducted many pointed observations of GX 339--4 and H1743--322 during their outbursts, with typical exposures $\sim$1--3 ks. 

Figs.\ \ref{lc}(a, b) show the light curves of GX 339--4 in the 1.5--12 keV range from the \xte All Sky Monitor (ASM; \citealt*{bradt93}) and the 15--50 keV from the Burst Alert Telescope (BAT; \citealt{barthelmy05,krimm13}) on board the \swift satellite, respectively. GX 339--4 underwent five major outburst during the $\sim$15 yr of the \xte operation, with transitions to the soft state in 1998--1999, 2002--2003, 2004--2005, 2007 and 2010--2011, see Fig.\ \ref{lc}(a). Fig. \ref{lc}(b) shows the \swift/BAT light curve from start of \swift operations in 2005 February until 2011 December. We see that GX 339--4 also underwent so-called `failed' outbursts in 2006 and 2009, during which it remained in the hard state. After the 2007 outburst, GX 339--4 brightened again for a few months, which we mark as the 2007 reflare. 

H1743--322 underwent a major outburst in 2003 and five minor outbursts in 2004, 2005 and 2009, as seen in the ASM light curve in Fig.\ \ref{h1743_lc}. We do not show the BAT light curve since the source was weak in the 15--50 keV band. Figs.\ \ref{lc}(a, b) and \ref{h1743_lc} also show the times of the pointed \xte observations of the sources.

\subsection{\textit{RXTE} pointed observations of GX 339--4 and H1743--322}
\label{obs_gx339}

Table 1 in \citet{corbel2013a} lists the radio fluxes at either 8.6 or 9.0 GHz from observations of GX 339--4 in the hard spectral state by the Australia Telescope Compact Array (ATCA) and corresponding \xte pointed observations. Hereafter, the hard spectral state is defined by the 3--10 keV unabsorbed spectral index of $\Gamma\la 2$, following \citet{remillard2006}. We have used standard spectral products\footnote{\url{heasarc.gsfc.nasa.gov/docs/xte/recipes/stdprod\_guide.html}} of the PCA (which include all available Proportional Counter Units and layers) and HEXTE data for the observations which were within a day from a radio pointing. However, we have not used four very low X-ray flux data, which were noisy. For the data starting 2005 December, we used only the HEXTE Cluster B because the Cluster A stopped working. On 2009 December 14, the Cluster B stopped working as well and for later data we used only the PCA. This yields a sample consisting of 70 quasi-simultaneous radio/X-ray observations over multiple outbursts, listed in Table \ref{GX}.

Table 1 in \citet{coriat2011} lists the radio fluxes at 8.5 GHz from observations of H1743--322 in the hard spectral state by the ATCA and the Very Large Array together with corresponding \xte pointed observations. We have utilised the standard spectral products for those observations in the same way as for GX 339--4. This yields a sample of 28 quasi-simultaneous radio/X-ray observations, listed in Table \ref{H}. 

\subsection{The Galactic Ridge emission}
\label{Ridge}

Both GX 339--4 and H1743--322 are located very close to the Galactic Plane and therefore the fluxes estimated from \xte have significant contamination from the Galactic Ridge emission \citep{valinia1998,revnivtsev2006}. To obtain the spectrum of that contribution for GX 339--4, we used the simultaneous \xte/PCA and \chandra observations taken during its quiescent state on MJD 52911. The \chandra spectra, given its high angular resolution, give the contribution from GX 339--4 alone, while the difference between the PCA and \chandra spectra gives the Ridge contribution to the PCA spectra. The \chandra spectra were fitted by an absorbed power law model, while those from the PCA by absorbed two power-law components, with one fixed at the parameters obtained from fitting the \chandra spectrum. In addition, a Gaussian line was added to model the Fe K fluorescent line. The obtained Galactic Ridge spectrum for the vicinity of GX 339--4 is in agreement with that obtained by \citet{revnivtsev2003}.

In the case of H1743--322, we followed \citet{kalemci2006} and \citet{coriat2011} and combined the eight \xte/PCA observations in the quiescent state made on MJD 53021--53055. We modelled the PCA spectrum with an absorbed power-law and an Fe K line, and obtained the spectral parameters consistent with those of \citet{kalemci2006} and \citet{revnivtsev2003}. We use this model to represent the Galactic Ridge spectrum for this source.  

\subsection{The X-ray spectral model and the bolometric fluxes}
\label{bolometric}

For spectral fitting, we used {\sc xspec} \citep{arnaud96}. For both sources, we modelled the primary continuum in the hard state as thermal Comptonization, using the {\tt nthcomp} model \citep*{zdziarski1996}. The low-energy photon spectral indices for all the observations were $\Gamma\la 2$, with the softest spectra being from H1743--322. When the data required it, we also used the {\tt ireflect} model \citep{magdziarz1995} to account for Compton reflection, with an added Gaussian line for the fluorescent Fe K emission. However, we found that accounting for reflection was statistically required only for high-flux spectra of GX 339--4. The seed photon temperature was kept at 0.2 keV in most cases. However, for some observations of H17443--322, we found an excess soft X-ray emission, which we modelled by the multicolour disc model {\tt diskbb}. In those cases, we assumed the disc inner temperature equals the seed photon temperature in the Comptonization component. We modelled the ISM absorption by the {\tt phabs} model. A multiplicative constant was used to account for a difference in the normalisation between the PCA and HEXTE. For all spectra, we included the respective Galactic Ridge components with the parameters fixed at those obtained in Section \ref{Ridge}. We fitted the PCA and HEXTE spectra in the energy ranges of 3--20 keV and 20--150 keV, respectively. 

For GX 339--4, we fitted the value of the ISM hydrogen column density, $N_{\rm H}$, except when it could not be constrained. For those cases, we fixed it at $6 \times 10^{21}$ cm$^{-2}$ \citep{zdziarski1998}. For H1743--322, there exist several estimates of $N_{\rm H}$; $1.6\times 10^{22}$ cm$^{-2}$ \citep{capitanio2009}, $1.8\times 10^{22}$ cm$^{-2}$ \citep{prat2009} and $2.3\times 10^{22}$ cm$^{-2}$ \citep{miller2006}. We have then fitted the \xte/PCA spectra of H1743--322 from the brightest outburst in 2003 \citep{miller2006}, and found $N_{\rm H}\simeq 1.7\times 10^{22}$ cm$^{-2}$, which we used for our observations with the simultaneous radio measurements. 

We have estimated the unabsorbed bolometric flux (excluding the Galactic Ridge emission) from the above model for all of the observations. Since the multi-wavelength spectra of all known BHXRBs are dominated by the X-ray band, we define the bolometric flux as that in the energy range of 0.1--$2 \times 10^3$ keV. We have estimated it using the {\tt cflux} model in {\sc xspec}. The largest uncertainty on the bolometric flux arises from the uncertainty in constraining the value of electron temperature, $kT_{\rm e}$, of the {\tt nthcomp} model. For observations with low X-ray fluxes, its value could not be constrained, and then we varied it in the range of 100--$10^3$ keV. We found that decreasing $kT_{\rm e}$ to below 100 keV led to unacceptable fits. At high X-ray flux observations, the values of $kT_{\rm e}$ are well constrained, and that allowed ranges contribute to the flux errors. For the high X-ray flux observations, the flux error was chosen to be the larger of the error estimated from {\tt cflux} and the flux uncertainty of the \xte/PCA, which has been estimated at 3 per cent by \citep{corbel2013a}. The reduced $\chi_{\nu}^{2} \sim 1$ for all the spectral fits. The photon index of the {\tt nthcomp} model was always $\Gamma \la 2$, confirming the sources were indeed in the hard state during the selected observations.

The lists of the radio and bolometric X-ray fluxes of GX 339--4 and H1743--322, along with their errors, are given in Tables \ref{GX} and \ref{H},  respectively. We denote the unabsorbed bolometric flux as $F_{\rm bol}$.

\begin{table}
\caption{The results of fitting a power law to the data. The errors correspond to the 90 per cent confidence ranges and 'f' denotes a fixed parameter. \label{fits}}
\centering
\begin{tabular}{c c c c}
\hline
Sample &     $a$      &     $b$      & $\chi^2/$d.o.f.     \\
\hline
GX 339--4       &  &      & \\
\hline
All     &     1.28 $\pm$ 0.01 &     0.806 $\pm$ 0.003 &  7798/68 \\
Rise ($b$ fixed) & 1.27 $\pm$ 0.01  &   0.806f &  6638/26 \\
Decline ($b$ fixed)  &  1.94 $\pm$ 0.01   &  0.806f       & 3226/34  \\
Rise    &     0.56 $\pm$ 0.01 &     1.050 $\pm$ 0.006 &  5832/25 \\
Decline  &     2.29 $\pm$ 0.03 &     0.62 $\pm$ 0.01  &  2500/33\\
\hline
H1743--322 &        &      &        \\
\hline
Rise         &     0.28 $\pm$ 0.01 &     1.20 $\pm$ 0.02  &  19441/9 \\
Decline   &     0.41 $\pm$ 0.03 &     0.19 $\pm$ 0.06  &  64/15   \\
\hline
\end{tabular}
\end{table}

\section{Results}
\label{results}

\subsection{The radio/bolometric flux correlations}
\label{correlations}

\begin{figure}
\includegraphics[height=\columnwidth,angle=-90]{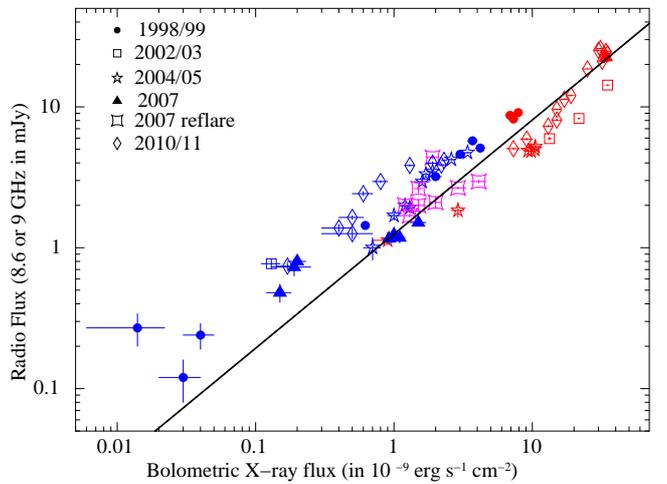}
\caption{The radio/bolometric flux correlation for GX 339--4 for the 1997--2011 pointed \xte observations. Different outbursts are marked with different symbols. The observations carried during the rise and decline of the hard state are shown in red and blue, respectively (following Fig.\ \ref{lc}), see Fig.\ \ref{cor} for separate plots. The magenta large squares correspond to the 2007 reflare, which assignment to either rise or decline appears uncertain. The solid black line shows the best fit of a single power law to the whole sample, with $b\simeq 0.81$.}
\label{cor1}
\end{figure}

\begin{figure*}
\includegraphics[scale=0.34,angle=-90]{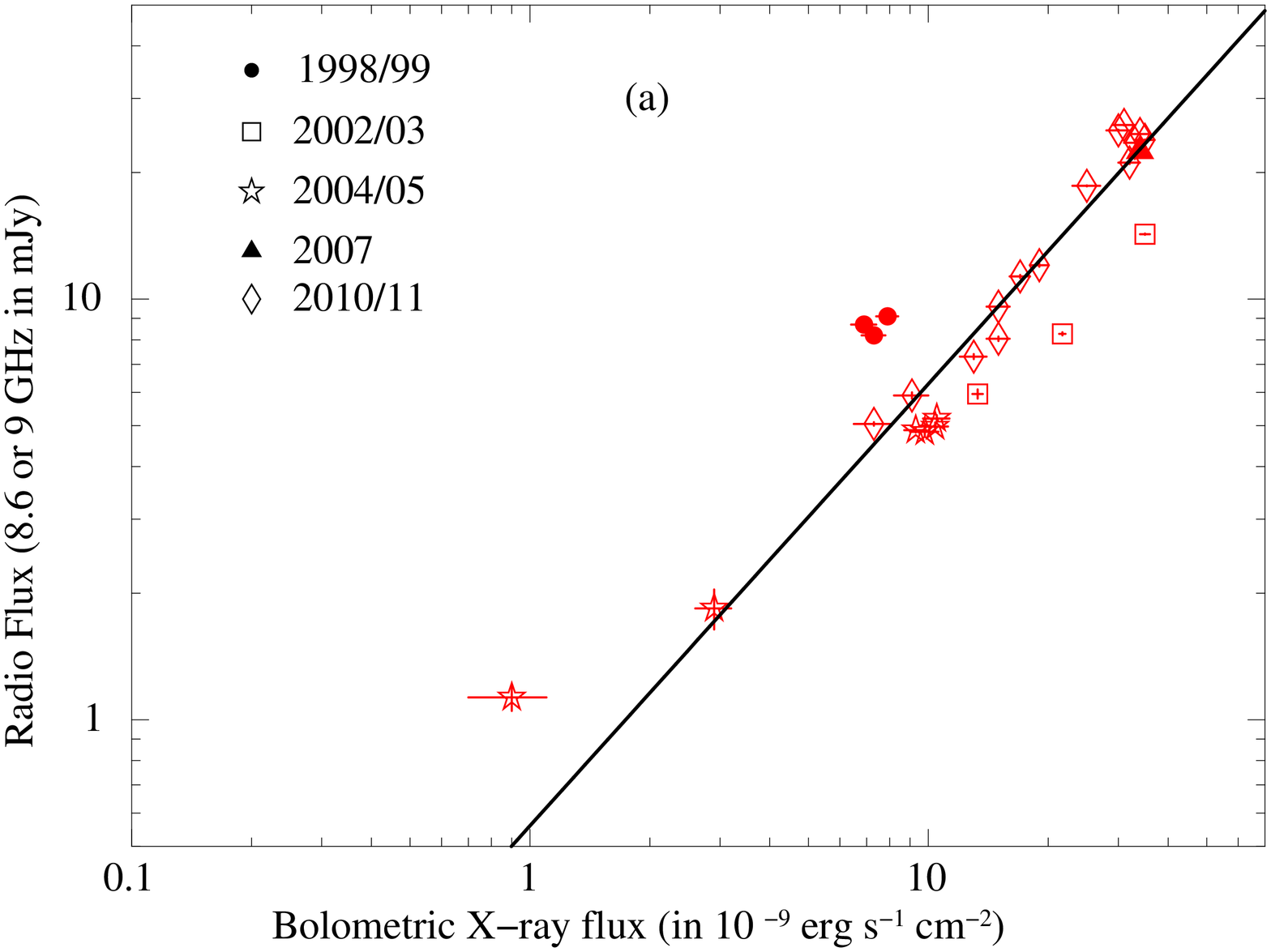}
\includegraphics[scale=0.34,angle=-90]{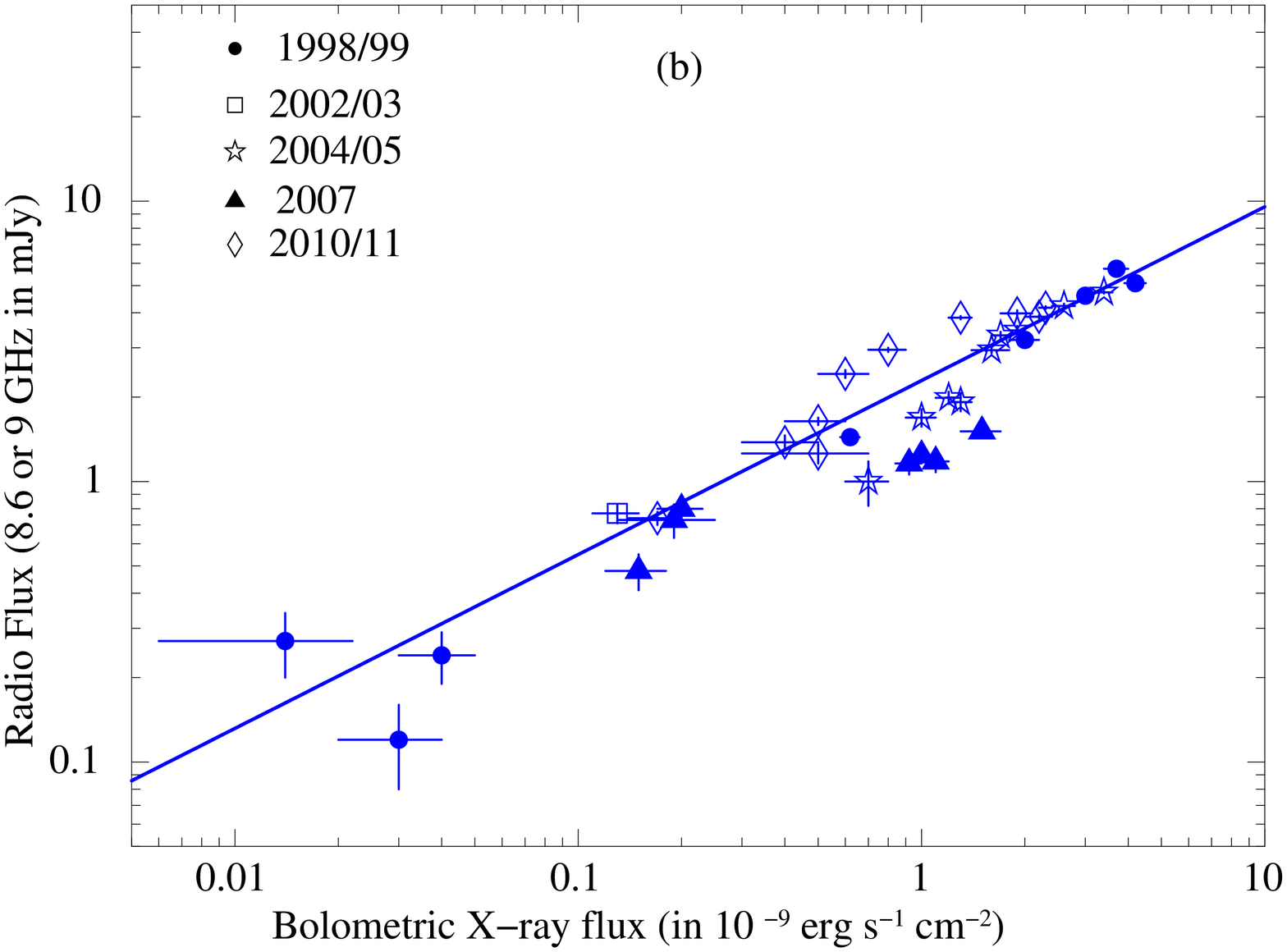}
\caption{The radio/bolometric flux correlation for GX 339--4 shown and fitted separately for (a) the rise and (b) the decline of outbursts. The symbols are the same as in Fig.\ \ref{cor1}. The 2007 reflare data points are not taken into account, see text. We clearly see that the data for the rise and decline follow different values of the fit index, with $b\simeq 1.1$ and $\simeq 0.6$, respectively. }
\label{cor}
\end{figure*}

\begin{figure}
\includegraphics[height=\columnwidth,angle=-90]{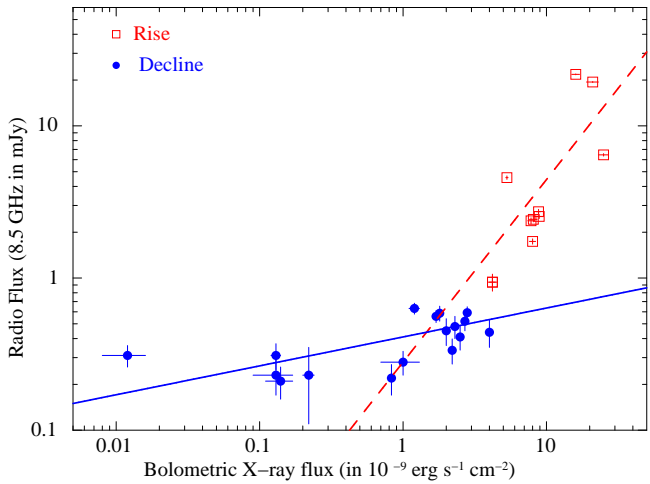}        
\caption{The radio/bolometric flux correlation for H1743--322. The observations carried during the rise and decline of the hard state are shown by red open squares and blue filled circles, respectively. The red dashed and blue solid lines show the best fit of a power-law to the rise and decline data, with $b\simeq 1.2$ and $\simeq 0.19$, respectively. }
\label{cor2}
\end{figure} 

\begin{figure}
\includegraphics[height=\columnwidth,angle=-90]{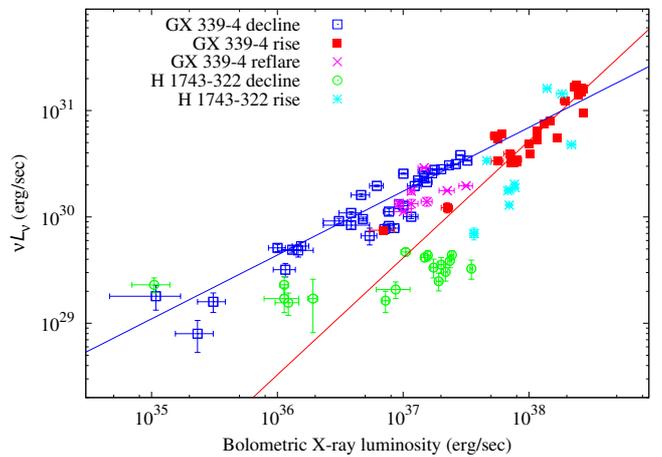}        
\caption{The radio versus unabsorbed bolometric luminosities for the hard state of both GX 339--4 and H1743--322, assuming $D=8$ and 8.5 kpc, respectively. We show the rise and decline by red filled squares and blue open circles for GX 339--4, and by cyan starred crosses and green open circles for H1743--322, respectively. Also, we show the 2007 reflare of GX 339--4 by magenta crosses. The solid lines represent the two fits to the GX 339--4 data. The vertical axis is for 8.6 GHz for GX 339--4 and 8.5 GHz for H1743--322.}
\label{cor3}
\end{figure} 

We show our results for GX 339--4 in Fig.\ \ref{cor1}, adopting the colour coding of the rise (red) and decline (blue) and the 2007 reflare (magenta) of Fig.\ \ref{lc}. We also identify different outbursts using different symbols. We first fit the entire data set with a power law, 
\begin{equation}
 \frac{F_{\rm R}}{\rm 1\,mJy} = a \left(\frac{F_{\rm bol}}{10^{-9}\, {\rm erg\, cm}^{-2}\, {\rm s}^{-1}}\right)^b,
\label{fit}
\end{equation}
symmetrically in $F_{\rm R}$ and $F_{\rm bol}$. We find the best-fit of $b\simeq 0.81$. The fit results are given in Table 3. Our best-fit power law is steeper than that found in previous radio/X-ray correlation studies in GX 339--4 \citep{corbel2003,corbel2013a,gallo2012}, who found $b\simeq 0.6$.

A significant part of the dispersion found in the correlation is attributed to the intrinsic variability by \citet{corbel2013a} (who used the 3--9 keV flux). Indeed, we see in Fig.\ \ref{cor1} that we see a significant difference between the rise and decline, and, as a secondary effect, different outbursts showing different behaviour. \citet{corbel2013a} attributed the former effect to a changing normalisation of the fitting power-law, but did not notice a significant difference in the correlation index. In order to test this hypothesis, we first fitted a power law separately to the rise and decline data keeping the index fixed at the value obtained for the entire data set, and then allowed a free index for each of the two sets. For those fits, we have removed the 2007 reflare data (8 observations), given that it is unclear whether those points should be classified as the tail of the decline of the main 2007 outburst or the rise of a new burst. Also, the 15--50 keV flux of the reflare showed two peaks, see Fig.\ \ref{lc}(b), which further complicates the event classification. Our results are given in Table \ref{fits} and Fig.\ \ref{cor}. We clearly find different indices for the rise and decline, $b\simeq 1.1$ and $\simeq 0.6$, respectively. 

{\it This hints at the possible existence of a hysteresis in the accretion-jet coupling}, in addition to the well-known hysteretic behaviour in the X-ray flux/X-ray hardness plane. We note that the flux/hardness hysteresis is seen only in the flux of the transition to and from the intermediate spectral state (connected then to soft states), see, e.g., \citet{dunn2008}. Within the proper hard state, both the rise and decline tracks are identical (for the overlapping flux ranges, i.e., below the return to the hard state for a given outburst). This suggests the same properties of the accretion flow in those flux ranges. Thus, the effect we have found points to different behaviour of the jet rather than the accretion flow. We see that the jet gets relatively brighter with decreasing luminosity with respect to the accretion flow luminosity (i.e., more radio loud) during the decline, while the radio loudness is roughly constant during the rise.

A caveat for this result is that there are almost no data during the rise for $F_{\rm bol}\la 5\times 10^{-9}$ erg cm$^{-2}$ s$^{-1}$, which is the range of the overlap between the rise and decline bolometric fluxes. There are in fact only two rise points below that $F_{\rm bol}$, with the lowest one at $\sim 10^{-9}$ erg cm$^{-2}$ s$^{-1}$, and only one of them departs from the decline track. This sparsity is apparently due to difficulty of quickly scheduling radio observations after the detection of an outburst. Therefore, we cannot be sure if we indeed see a hysteresis (which is the existence of two possible states for the same range of a parameter, $F_{\rm bol}$ in our case), or just a change in the shape of the correlation. An additional complication is that we see somewhat different tracks for individual outbursts. 

A similar effect has been found the near infrared (the H band) vs.\ X-ray flux correlation in the transient BHXRB XTE J1550--564 \citep{russell2007}. Its rise track, when extrapolated to low fluxes, lies significantly below the decline track, see fig.\ 2 of \citet{russell2007}. On the other hand, no difference was noted in the infrared/X-ray flux correlation for GX 339--4 between the rise and decay of outbursts \citep{coriat2009}, and those authors noted only a break in that correlation for GX 339--4 at low fluxes. We cannot constrain the presence of a break in the radio/X-ray flux correlation of GX 339--4 due to the sparsity of the data at low fluxes.  

We show our results for H1743--322 in Fig.\ \ref{cor2}, following our adopted colour coding of the rise (red) and decline (blue). We can see very different tracks for the rise (at high fluxes only) and decline (at low fluxes). This is similar to the results of \citet{coriat2011}. We then fit a power-law dependence, equation (\ref{fit}), to the rise and decline data separately. We show the fit results in Table \ref{fits}, and plot them in Fig.\ \ref{cor2}. We see that the rise and decline indices are markedly different, $b\simeq 1.2$ and $\simeq 0.19$, respectively. While those values of $b$ are qualitatively similar, they are quantitatively different from either the standard, $b\sim 0.6$ or outlier, $b\sim 1.4$ power-law indices reported earlier \citep{corbel2013a}. Our fit index for the decline is similar to that of \citet{jonker2010} obtained for the low luminosity data of the 2008a outburst. We see that the correlation index for the rise is similar to that of the corresponding one in GX 339--4. However, the decline in H1743--322 clearly follows a different track than that of GX 339--4. Given almost no overlap between the rise and decline tracks in H1743--322, we do not know if there is a jet hysteresis or just a change of the slope of the correlation at $F_{\rm bol}\sim 5\times 10^{-9}$ erg cm$^{-2}$ s$^{-1}$, similarly to the case of GX 339--4.

\subsection{Comparison of the correlation in different sources}
\label{comparison}

\begin{figure}
      \includegraphics[width=\columnwidth]{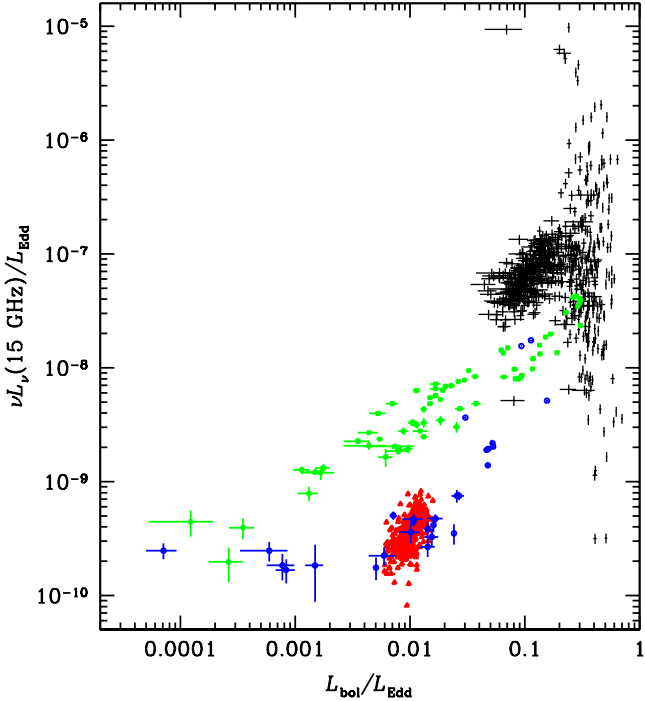}        
\caption{The correlation between the radio 15 GHz luminosity, $\nu L_\nu$, versus the unabsorbed bolometric X-ray luminosity, both in units of the Eddington luminosity, $L_{\rm Edd}$ [$\equiv 1.25\times 10^{38}\times 2/(1+X)$ erg s$^{-1}$, where $X$ is the H abundance], for Cyg X-3 in all states (black error bars), and Cyg X-1 (red triangles with error bars), GX 339--4 (green squares with error bars) and H1743--322 (blue circles with error bars), all three in the hard state. For GX 339--4 and H1743--322, we assumed our default $M_{\rm BH}$ and $D$ and $X=0.7$. For Cyg X-1 and Cyg X-3, we assumed $M_{\rm BH}=15$, $5\msun$, $D=1.86$, 7 kpc, $X=0.7$, 0, respectively.
}
\label{cor4}
\end{figure} 

In Fig.\ \ref{cor3}, we plot the radio versus bolometric luminosities for GX 339--4 and H1743--322 together. \citet{coriat2011} showed that the steep correlation of H1743--322 at high luminosities with their value of $b \simeq 1.4$ connected to the high luminosity correlation for GX 339--4 with $b\simeq 0.6$. In our work, we also find the connection but $b \simeq 1.1$ for the rise of {\it both\/} GX 339--4 and H1743--322. On the other hand, the decline tracks are different for H1743--322 and GX 339--4, and they intersect at a low luminosity of $\sim\! 10^{35}$ erg s$^{-1}$. \citet{coriat2011} proposed that the steep index of $b\simeq 1.4$ corresponded to a high efficiency of accretion, while $b\simeq 0.6$ corresponded to a radiatively inefficient accretion. In the light of our present results, it appears that both the rise tracks of GX 339--4 and H1743--322, established for high fluxes only, correspond to accretion being much more efficient than that for the decline, corresponding to low luminosities. This is in agreement with the fact that the bolometric luminosity during hard-to-soft state transition changes by at most a factor of two, which argues against the luminous hard state being radiatively inefficient. On the other hand, while both of the decline tracks may correspond to radiatively inefficient accretion, the cause of the difference in the decline indices between GX 339--4 and H1743--322 remains unclear.  

Fig.\ \ref{cor4} shows a plot of 15 GHz radio luminosity versus the unabsorbed bolometric X-ray luminosity, both in the units of Eddington luminosity, $L_{\rm Edd}$, for GX 339--4, H1743--322, and for Cyg X-1 (from \citealt{zdziarski2011}) and Cyg X-3 (from \citealt{zdziarski2016}). The 15 GHz fluxes for GX 339--4 and H1743--322 have been extrapolated from 8.5--9.0 GHz to 15 GHz assuming the radio spectral indices ($F_\nu\propto \nu^\alpha$) of $\alpha=0.4$ \citep{corbel2013a} and 0 \citep{coriat2011} for GX 339--4 and H1743--322, respectively. 

Fig.\ \ref{cor4} suggests a somewhat different picture than Fig.\ \ref{cor3}. In terms of the Eddington ratio, the tracks for GX 339--4 and H1743--322 look quite different. GX 339--4 shows an overall single slope, while H1743--322 does have a pronounced break at $L_{\rm bol}/L_{\rm Edd}\simeq 0.01$. The hard state of Cyg X-1 lies close to that break, and the slope of its correlation is roughly similar to that H1743--322. The correlation in the hard state of Cyg X-3 is somewhat fuzzy, but nevertheless its slope appears similar to that of GX 339--4. The two overlap within about a decade, but Cyg X-3 is more radio loud, by a factor of $\sim$2--3. It is possible that the high radio loudness of Cyg X-3 is due to interaction of its jet with the strong wind of its Wolf-Rayet companion, while no such effect appears in Cyg X-1 due to a different value of the ratio between the jet and wind powers, see \citet*{yoon16}. We shall note significant uncertainties in the normalization of the Eddington ratios due to the uncertainties in both the distances and masses of the objects. Furthermore, the radio emission in high mass X-ray binaries can be substantially absorbed in the stellar wind, see \citet{zdziarski12} for a study of its effect in Cyg X-1. 

We finally note while we have considered integrated energy fluxes in the X-ray range, our radio fluxes are those measured at single frequencies, at 8.6 or 9 GHz in the case of GX 339--4, and at 8.5 GHz in the case of H1743--322. In Fig.\ \ref{cor4}, which compares our results to those for two high-mass X-ray binaries, we have also used extrapolations of those fluxes to 15 GHz based on the observed radio slopes in the hard state (which show generally flat radio spectra). Making an assumption that the uncertainty on $\alpha$ is 0.2 \citep{espinasse2018}, the error introduced in that way is small, $<$12 per cent. Then, using single-frequency radio flux measurements in the case of $\alpha\simeq 0$ allows a determination of most of the properties of partially synchrotron self-absorbed jets, as shown first by \citet{blandford1979}, and then, e.g., by \citet{heinz2003}. However, while the radio indices in the hard state are generally flat, they are still often different from null. In fact, \citet{espinasse2018} have shown that there is a difference in the radio spectral indices of the sources lying on the standard track and the outlier sources. A future analysis of the correlation taking into account also the radio spectral index is then desirable. It is, however, outside the scope of this work since it would involve one more dimension of the analysis ($\alpha$) and require the use of a different data set.

\section{Conclusions}
\label{conclusions}

We have investigated the correlations of the radio flux with the bolometric flux in the hard state of two low-mass BHXRBs, GX 339--4, the main representative of the standard radio/X-ray correlation, and H1743--322, the main outlier source. Our main results are as follows.

Based on the \xte PCA and HEXTE data, we have calculated the bolometric fluxes (which are dominated by the X-rays) for $\sim$100 hard-state observations of GX 339--4 and H1743-322 quasi-simultaneous with radio observations.

We have calculated the indices of the correlations the radio versus the bolometric flux. In the case of GX 339--4, we found $b\simeq 0.8$ when all the data are fitted, which is significantly different from $b\simeq 0.6$ found when the narrow-band 3--9 keV fluxes were used. The data for different outbursts show, however, deviations from the single power-law correlation, as shown in Fig.\ \ref{cor1}. We have found a significant difference between the slopes of the correlation when we fit separately the data for the outburst rise and decline, namely $b\simeq 1.1$ and $\simeq 0.6$, respectively. 

In the case of H1743--322, we have found $b\simeq 1.2$ for the rise, i.e., a similar value as that for the rise of GX 339--4. This indicates the similarity in the jet/accretion flow coupling in both sources during the rise and at high luminosities (for which the data are available). On the other hand, the decline index of H1743--322 is $\simeq 0.2$, which is significantly different from $\simeq$0.6 in GX 339--4.

Our results hint at a possible existence of jet hysteresis in those sources, namely different jet radio luminosities for the same bolometric (dominated by the accretion-flow) luminosity depending on the source history, i.e., either an outburst rise or its decline. The presence of hysteresis would be clearly established if there are more radio data at low X-ray luminosities during the rise of an outburst. Presently, the available rise data correspond almost exclusively to high luminosities, while the decline data are for low luminosities only. The latter is due to the well-known X-ray hysteresis, i.e., the soft-to-hard transitions in transients occurring at much lower luminosities than the hard-to-soft ones. In the case of H1743--322, the shape of the radio correlation changes at $L_{\rm bol}\sim 0.01 L_{\rm Edd}$.

We have compared the correlations in GX 339--4 and H1743--322 with those in the high-mass BHXRBs Cyg X-1 and Cyg X-3, see Fig.\ \ref{cor4}. We have found that the correlation in Cyg X-1 appears similar to that of H1743--322. The hard state of Cyg X-3 is then similar to the luminous hard states of GX 339--4. The latter correlations have approximately the same slope, while Cyg X-3 is more radio loud by a factor of $\sim$2--3.

\section*{Acknowledgements}
This research has been supported in part by the Polish National Science Centre grants 2013/10/M/ST9/00729 and 2015/18/A/ST9/00746, and it has made use of data obtained from the High Energy Astrophysics Science Archive Research Center (HEASARC), a service of the Astrophysics Science Division at NASA/GSFC and of the Smithsonian Astrophysical Observatory's High Energy Astrophysics Division. We thank the anonymous referee for insightful comments.
 
\bibliography{bibtex}{}
\bibliographystyle{mnras}

\label{lastpage}
\end{document}